\documentclass[aps,prl,twocolumn,groupedaddress,showpacs]{revtex4}
\usepackage{amsmath}
\usepackage{graphicx}

\begin{document}


\title{Confinement without boundaries:\\ Anisotropic diffusion on the surface of a cylinder}


\author{Remy Kusters, Stefan Paquay and Cornelis Storm}
\affiliation{Department of Applied Physics and
Institute for Complex Molecular Systems, Eindhoven University of
Technology, P. O. Box 513, NL-5600 MB Eindhoven, The Netherlands}


\date{\today}

\begin{abstract}

Densely packed systems of thermal particles in curved geometries are frequently encountered in biological and microfluidic systems. In 2D systems, at sufficiently high surface coverage, diffusive motion is widely known to be strongly affected by physical confinement, e.g., by the walls. In this Letter, we explore the effects of confinement by shape, not rigid boundaries, on the diffusion of particles by confining them to the surface of a cylinder. We find that both the magnitude and the directionality of lateral diffusion is strongly influenced by the radius of the cylinder. An anisotropy between diffusion in the longitudinal and circumferential direction of the cylinder develops. We demonstrate that the origin of this effect lies in the fact that screw-like packings of mono- and oligodisperse discs on the surface of a cylinder induce preferential collective motions in the circumferential direction, but also show that even in polydisperse systems lacking such order an intrinsic finite size confinement effect increases diffusivity in the circumferential direction.

\end{abstract}


\maketitle

The influence of confinement on the dynamics of glassy systems has been the focus of numerous studies on particles and polymers confined to narrow channels, or between parallel plates \cite{Varnik2002, Batistakis2012, Lucena2012}. These works showed that the glass transition exhibits non-monotonic behavior, depending sensitively on particle-wall interactions \cite{Mittal2008, Lang2010, Mandal2014}. In this Letter, we seek to eliminate the interactions with walls to achieve a purely geometrical confinement. We do so by studying 2D diffusive processes that take place on the  surface of a cylinder. This way confinement without boundaries is achieved. We map out how the diffusive motion of discs on this space is affected by the geometry of the confinement. 

The system we consider is far from academic: highly curved \cite{Shum2010, Irvine2010, Ershov2013, Kusters2013, Kusters2014} and crowded motifs (see, e.g., \cite{Hofling2013}  and references therein) are abundant in biological and microfluidic systems. In these systems crowding and shape significantly contribute to the effective diffusivity of particles which, in the case of membrane-associated proteins, is completely confined to a curved, 2D substrate. Likewise, rapidly evolving microfuidic techniques such as topological emulsions \cite{Shum2010} and surface-confined colloids \cite{Ershov2013}  yield systems where crowding and curvature meet directly. Finally, the dynamics of particles in highly crowded and confined environments has been the subject of numerous studies (experimental and theoretical) in the context of flowing glassy materials, for instance in the rheological behavior of jammed emulsions in microchannels \cite{Goyon2008, Goyon2010, Bocquet2009,Paredes2013}.

The effects of crowding and shape, individually, have been extensively characterized in the past: experimental and theoretical work in the context of membrane-bound diffusion have clearly established the generic effects of shape on diffusive timescales \cite{Ashby2006, Schuss2007, Holcman2011, Kusters2013, Kusters2014}, the upshot being that shape, in itself, may alter and even direct diffusion in dilute particulate systems. The effects of pure crowding on the diffusive properties in two and three dimensions, likewise, has been widely studied (see, e.g., \cite{Hofling2013}  and references therein). 

\begin{figure}
\centering
\includegraphics[scale=0.55]{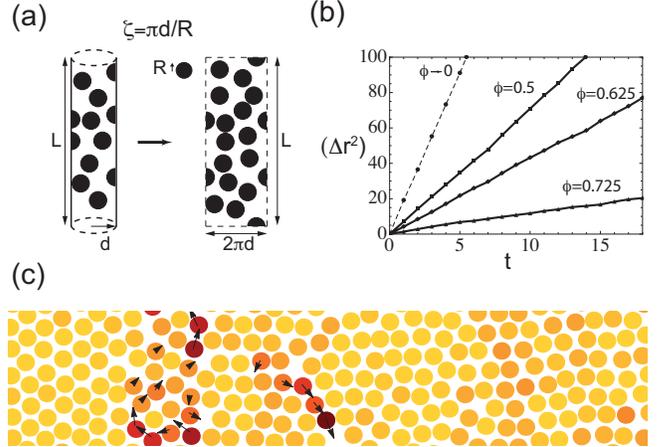}
\caption{ (a) The system we consider: A cylinder with length $L$, circumference 2 $\pi \times d$ covered with particles with radius $R$ that occupy a fraction $\phi$ of the cylinder's area. The ratio between the cylinder's circumference and the particle size we call $\zeta = \pi \times d/R$. (b) The MSD $\langle\Delta r^2\rangle$. for various surface coverages ($\phi =$ 0, 0.5, 0.625 and 0.725, and $\zeta = 1.7$). In this regime, the motion is diffusive; $\langle\Delta r^2\rangle = 4 D(\phi) t$. In (c), red colors label the most mobile particles, and arrows indicate the cumulative displacement vector of individual particles after 10000 steps (vector shown only for particles that have moved at least twice the system mean). In these crowded cylinder packings, the long-time diffusivity is dominated by cage rearrangements: the coherent motion of clusters of particles.\label{fig:0}}
\end{figure} 

In the present work, we study the combined effects of crowding and confinement on the diffusion of discs on the surface of a narrow cylinder. Recent works have shown that at high surface coverages, the packing of circles on a cylinder is elaborately organized into various screw-like packings, where the symmetry and the tilting axes of the screw structure depend on the ratio between the radius of the crowders, $R$, and the radius of the cylinder, $d$, \cite{Mughal2011, Mughal2013, Mughal2014}. While this was established for athermal (granular) packings, we now measure whether these naturally arising screw-like structure alter the diffusive motion of the particles, particularly if $d$ is not significantly larger than $R$. For ease of viewing, we unfold the cylinder to yield a flat, two-dimensional sheet with periodic circumferential boundaries. We also take the longitudinal direction to be periodic to avoid boundary effects in that direction, but will choose lengths $L$ to be much larger than either $R$ or $d$  (See Fig. \ref{fig:0} (a)). In the following, we denote by $\zeta$ the ratio of the circumference of the cylinder to twice the radius of the diffusing particles (which we fix at $R=0.5$):  $\zeta = \pi d / R$.

Using the molecular dynamics package LAMMPS \cite{plimpton1995}, we execute Langevin dynamics simulations. To prevent unphysically large overlap between particles in the initial configuration the following scheme is used to prepare the system: First, a random configuration of discs on the surface is generated, after which we equilibrate  with a soft pair potential of the form $V_{ij} = V_0\left( 1 + \cos(\pi r_{ij}/r_c)\right).$ $V_0$ is the maximum of the potential, $r_{ij}$ the distance between particles $i$ and $j$ and $r_c$ is the cutoff distance. During this run the constant $V_0$ is linearly ramped up from $V_0 = 0$ to $V_{max} = 3000 k_BT$). This ensures that initially overlapping particles are smoothly repelled until they no longer overlap, while any excess kinetic energy incurred by the forces is dissipated due to the viscosity term in the Langevin equation. After this equilibration run, the smooth potential is replaced by the repulsive part of a 6-12 Lennard-Jonnes potential, where we have chosen to cut off the potential at one Lennard-Jones length unit. We have verified that the results presented are insensitive to both the exact form of the repulsive potential, and to the cutoff distance of the potential itself (See {\em Supp. Fig. I}). 

\begin{figure}
\centering
\includegraphics[scale=0.48]{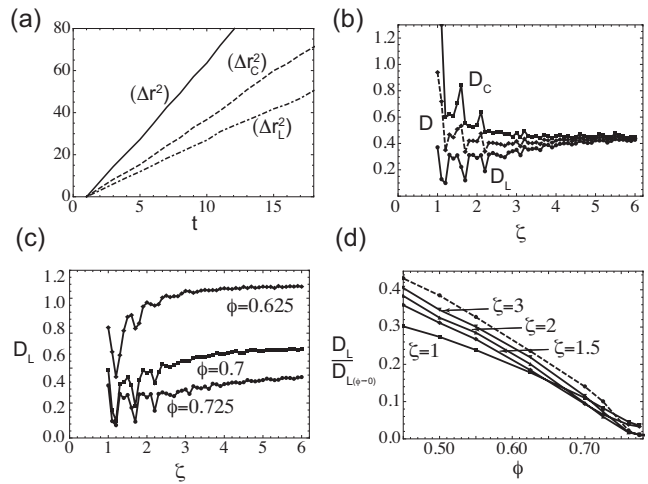}
\caption{ (a) Time evolution of the total Mean Squared Displacement $\langle\Delta r^2\rangle$ and its components in the circumferential direction $\langle\Delta r_C^2\rangle$ and the longitudinal direction $\langle\Delta r_L^2\rangle$ ($\zeta = 1.9$, $\phi = 0.625$ and $N=2500$). (b) shows the time evolution of the longitudinal ($D_L$), circumferential ($D_C$) and total ($D$) diffusivities as function of $\zeta$ ($\phi=0.725$ and $N=2900$). (c) The longitudinal diffusivity $D_L$ as function of $\zeta$ for various surface coverages ($\phi =$ 0.625, 0.7 and 0.725 and corresponding $N = $ 2500, 2700 and 2900). (d) The longitudinal component of the diffusivity $D_L$, relative to its value for $\phi \rightarrow 0$, as a function of the surface coverage $\phi$ for $\zeta =$ 1, 1.5, 2 and 3 (solid lines) and for a square $L=2 \pi d$ (Dashed line). The total system size was kept constant at $2 \pi d \times L = 1570.8$, and $\phi$ was varied by varying $N$, the number of particles in the system. \label{fig:1}}
\end{figure}

We focus on the regime where the Mean Squared Displacement (MSD) scales linear with time, $\langle\Delta r^2\rangle= 4 D(\phi) t$ (See Fig. \ref{fig:0} (b)). In highly crowded systems, the long-term diffusivity is dominated by global rearrangements of large clusters of particles, caused by so called cage-rearrangements \cite{Donati1999, Weeks2002, Narumi2011}. In Fig. \ref{fig:0} (c) we illustrate the displacement of particles over the course of 10000 timesteps, and indicate with arrows the particles that have traveled more than twice the average displacement $\Delta r > 2 \Delta r_{MSD}$. In the narrow systems we consider, there is a natural distinction between the longitudinal and circumferential direction: the longitudinal dimension is significantly larger than the circumferential one. In Fig. \ref{fig:1} (a), we decompose the total MSD into a longitudinal and  circumferential component: $\langle\Delta r^2\rangle= \langle\Delta r_C^2\rangle+ \langle\Delta r_L^2\rangle$. This reveals that, at least for this system ($\phi = 0.625$ and $\zeta = 1.9$), the effective displacement in the longitudinal direction is significantly smaller than the circumferential displacement (See Fig. \ref{fig:1} (a)).

\begin{figure*}
\centering
\includegraphics[scale=0.62]{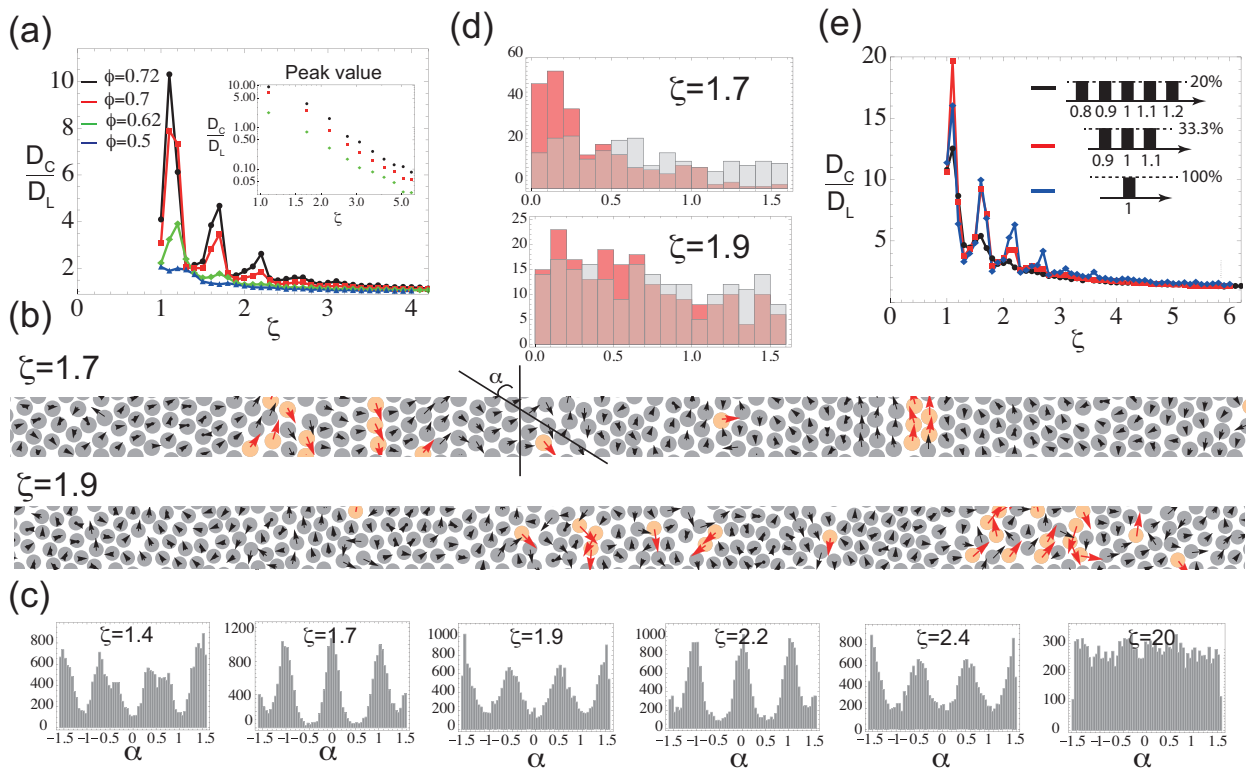}
\caption{(a) The ratio of the circumferential and lateral diffusivities as function of $\zeta = \pi d / R$ for various surface coverages $\phi$. The inset shows that the peak values show a powerlaw dependence on $\zeta$, though we have not yet established this over a significant range of $\zeta$s. Fig (b) shows a typical image of the system for $\zeta  = 1.7$ and $\zeta = 1.9$, indicating two distict orientations (These values of $\zeta$ were chosen to illustrate the behavior in adjacent maxima ($\zeta  = 1.7$) and minima $\zeta  = 1.9$ in the ratio $D_C/D_L$). Arrows indicate the dispacement of the particles for $\Delta t = 10000$, where black arrows indicate those displacements with $\Delta r <  2\times \Delta r_{MSD}$, and red arrows those with $\Delta r > 2\times \Delta r_{MSD}$. Fig. (c) shows the preferential bond angles $\alpha$, indicating the distribution of relative orientations of the six nearest neighbor pairs ($\phi = 0.725$ and $N=2900$). In (d) we separate out the orientation of the displacements relative to the circumferential directions for the red (mobile) particles and the gray (immobile) particles. This demonstrates, that although the overal motion of the slow particles is invariably isotropic, those particles with the highest diffusivity for $\zeta  = 1.7$ selectively move in the circumferential direction (small angles), whereas at $\zeta  = 1.9$ they, too, move isotropically (e) The effect of polydispersity on the extent of the anisotropy in $D_C/D_L$, for $\phi= 0.75$. We consider a monodisperse system (blue curve), a tridisperse system with an equal amount of particles with a radius of 0.9, 1 and 1.1 (red curve) and a polydisperse system with particle radii 0.8, 0.9, 1.0, 1.1 and 1.2 (black curve).   \label{fig:2}}
\end{figure*} 

 In the following, we demonstrate that this geometrically induced anisotropy is generic in these systems. We determine the $\zeta$-dependence of the longitudinal diffusivity, $D_L = \langle\Delta r_L^2\rangle/ 2 t$, the circumferential diffusivity, $D_C = \langle\Delta r_C^2\rangle/2 t$, and the total diffusivity, $D =  D_L + D_C$, for a fixed surface coverage $\phi =0.725$ (See Fig. \ref{fig:1} (b)). We find that upon increasing the radius of the system, and thus the number of particles that will fit the circumference of the cylinder, the longitudinal diffusivity increases while the circumferential component decreases. For $\zeta \gg 5$ the two eventually become equal as the system tends to a homogeneously flat, (2D) crowded system. $D_L$ and $D_C$ are not monotonic functions, they show distinct minima and maxima. The total diffusivity $D$ also shows a non-monotonic dependence on $\zeta$, not unlike the anisotropy observed in wall confined systems \cite{Lang2010, Mandal2014}.  Next,  Fig. \ref{fig:1} (c) shows that for various values of the surface coverage, the anisotropy as function of $\zeta$ remains. In Fig. \ref{fig:1} (d), we show the decrease in longitudinal component $D_L$, relative to its value for $\phi \rightarrow 0$, $D_{L(\phi \rightarrow 0)}$, as function of $\zeta$, for various surface coverages $\phi$ and shows that the trend observed in Fig. \ref{fig:1} (a),(b) and (c), i.e., an increasing longitudinal diffusivity upon increasing $\zeta$, indeed appears for a broad range of surface coverages $\phi$. The dashed line corresponds to a system whose length is equal to the circumference $L = 2 \pi d$ and thus exhibits isotropic diffusion. Eventually at $\phi \approx 0.82$ the diffusivity vanishes as the system becomes glassy. 

In Fig. \ref{fig:2} (a) we calculate the ratio $D_C/D_L$ as a function of $\zeta$ for various surface coverages $\phi$, and find that although it shows an overall decrease with increasing $\zeta$, maxima in the anisotropy arise periodically. The period of these peaks equal $\zeta = 0.5$ and they occur each time one additional particle precisely fits the circumference of the cylinder. The height of these peaks in $D_C/D_L$ decreases as a power law with an  exponent of approximately 3, though we have a limited range and cannot firmly establish this scaling. We note, too, that the magnitude of the heterogeneity decreases with decreasing $\phi$, establishing that it is truly the {\em combination} of shape and crowding, and not just the shape itself, that causes the effects we describe here. 

The question we address now is whether the observed effect is a consequence of the configurational order of the particles, i.e., of the screw-like patterns that naturally arise in these dense systems \cite{Mughal2011, Mughal2013, Mughal2014}. To quantify this, we apply the standard Delaunay triangulation scheme to find the lines connecting each particle to its 6 nearest neighbors. Then, we calculate the orientation (relative to the the circumferential axis, see Fig. \ref{fig:2} (b)) of these lines. To properly account for the bonds at the periodic edges of the system we also include bonds with the mirror images of the system.  Next, we tabulate the distribution of these angles $\alpha$ for $\phi = 0.725$ at those $\zeta$ where $D_C/D_L$ exhibits extrema (see Fig. \ref{fig:2} (c)). We see that the anisotropy in diffusivity is strongly correlated to the spatial configuration of the particles. Specifically, we find that the structures where the dominant orientation is in the longitudinal direction, which correspond to $\alpha = 0$ and $\alpha =1.1$ as peak values, have a larger diffusivity in the circumferential direction compared to systems which have $\alpha = 0$ and $\alpha = 0.5$. These two states would, if densely packed, correspond to the symmetric packing and its affinely rotated structure discussed by Mughal {\em et al}. \cite{Mughal2014}.  

Our observations strongly suggests that the configurational order of the system causes the diffusive anisotropy. The long-term diffusivity in highly crowded systems is generally dominated by so called cage-rearrangements \cite{Donati1999, Weeks2002, Narumi2011}. We hypothesize that the cage rearrangements in these screw-like configurations induce motion in a preferred direction. To verify this, we have calculated the orientation and the magnitude of the displacement for various values of $\zeta$. As may be seen in Fig. \ref{fig:2} (b), those particles with a displacement larger than twice the mean, $\Delta r > 2\times \Delta r_{MSD}$ (indicated with red arrows), appear in clusters. If we now visualize the motions of these clusters, we find that for $\zeta = 1.7$, their motion and thus the cage rearrangements, promote motion in the circumferential direction, while this motion is not directed for $\zeta = 1.9$. To further quantify this, we plot in Fig. \ref{fig:2} (d) the absolute value of the angle of the displacement vectors, relative to the y-axes for red (mobile) particles, which have $\Delta r > 2\times \Delta r_{MSD}$, and grey (immobile) particles, which have $\Delta r < 2\times \Delta r_{MSD}$. Particularly for $\zeta=1.7$, where we previously saw that the anisotropy is maximal, the slow particles do not show any directionality but the the fast moving clusters - that dominate the diffusivity - markedly do so. 

Based on previous work on dense suspensions under confinement, we also expect to see some intrinsic effects of confinement as a result of a confinement dependent flow cooperativity, known to affect the diffusivity in narrow channels \cite{Goyon2008, Goyon2010, Bocquet2009,Paredes2013}. In order to establish whether this is also at play here, we suppress the tendency to order by introducing polydispersity into the system. In Fig. \ref{fig:2} (e) we plot the ratio $D_C/D_L$ as function of $\zeta$ for various degrees of polydispersity (blue curve: monodisperse, red curve: tridisperse (0.9, 1.0, 1.1) and the black curve where we consider five different particle diameters (0.8, 0.9, 1.0 1.1, 1.2 )). We find that for increasing polydispersity the peaks in $D_C/D_L$ are reduced, confirming the configurational nature of the anisotropy, but that the increasing trend for decreasing $\zeta$ remains. We speculate, that this may be related to a side effect of wall confinement in glassy flows, where fragile zones appear to live much longer near the wall than they do in the bulk \cite{Goyon2008, Goyon2010, Bocquet2009,Paredes2013}. Even absent a physical wall, our systems are confined in the circumferential direction and we expect collective rearrangements in that direction to live longer than those along the long axis of the system.

To summarize, we have addressed the question of how the lateral diffusion of densely packed particles, confined to a cylinder, is affected by the shape of the substrate and the packing fraction. We find that the shape of the substrate in combination with crowding induces an anisotropy between the longitudinal and the circumferential diffusivity. We find that the extrema of the anisotropy coincide with an ordered packing of discs in a triangular lattice and its affinely rotated structure, geometries also observed for dense athermal packings \cite{Mughal2014}. These skewed orderings promotes collective motion in the circumferential direction. 

Our work reveals that diffusive motion along the surface of a narrow, crowded tube is hindered by an interplay between local geometry and crowding. Diffusivity in highly crowded systems is dominated by the collective motion of clusters and the size of these clusters may, especially at high coverage, become comparable to the local radii of curvature of the system. Though we consider only cylinders here, diffusion of particles that itself are relatively small compared to the local radius of curvature may very well be affected by the curvature of their substrate; not by their proper size, but by that of the cluster in which they diffuse, which can be comparable in size to the local radius of curvature.  Some specific biological settings where dense packings as well as high curvatures feature are cristea in mitochondria, dendritic spines in synapses and grana thylakoids in choroplasts, where the sizes of narrow membrane tubes, connecting these microdomains to the remainder of the cell, are not that much larger than the protein complexes that diﬀuse on it. From a practical viewpoint, engineered structures such as colloids confined to an interface \cite{Ershov2013} and topological emulsions \cite{Shum2010} are systems where curvature and crowding directly meet and would be ideal candidates to experimentally verify this geometrically induced anisotropic diffusion. 

We thank Thijs Michels and Claus Heussinger for valuable discussions. This work was supported by funds from the Netherlands Organization for Scientific Research (NWO-FOM) within the programme "Barriers in the Brain: the Molecular Physics of Learning and Memory" (No. FOM-E1012M) and grant number RGP0017/2012 awarded by the Human Frontier Science Program.
\bibliography{biblio}

\end{document}